\documentclass[reprint,onecolumn,amsmath,amssymb,jap,aip]{revtex4-2}
\usepackage{graphicx}
\usepackage{dcolumn}
\usepackage{bm}
\usepackage{epsfig}
\usepackage{mathptmx}
\usepackage{times}
\usepackage{amsmath}
\usepackage{amssymb}
\usepackage{dcolumn}
\usepackage{units}
\usepackage{upgreek}
\usepackage{tabularx}
\usepackage{todonotes}
\usepackage{threeparttable}
\usepackage[space]{grffile}
\usepackage[normalem]{ulem}

\usepackage{subfig}
\usepackage{multirow}
\usepackage{array}
\usepackage{soul}
\usepackage{svg}
\usepackage{bbm}
\usepackage{verbatim}
\usepackage{multirow}
\usepackage{booktabs}
\usepackage{mathtools}
\usepackage{mismath}

\usepackage{xspace}

\DeclareCaptionLabelSeparator{bar}{\space\textbar\space}
\renewcommand{\figurename}{Fig.}
\usepackage[font={small},]{caption}
\newcolumntype{C}{>{$}c<{$}}
\AtBeginDocument{
\heavyrulewidth=.08em
\lightrulewidth=.05em
\cmidrulewidth=.03em
\belowrulesep=.65ex
\belowbottomsep=0pt
\aboverulesep=.4ex
\abovetopsep=0pt
\cmidrulesep=\doublerulesep
\cmidrulekern=.5em
\defaultaddspace=.5em
}

\begin{document}
\title{In-memory factorization of holographic perceptual representations}
\author{Jovin Langenegger} \affiliation{IBM Research -- Zurich, S\"{a}umerstrasse 4, 8803 R\"{u}schlikon, Switzerland.}\affiliation{Department of Information Technology and Electrical Engineering, ETH Z\"{u}rich, Gloriastrasse 35, 8092 Z\"{u}rich, Switzerland.}

\author{Geethan Karunaratne} \affiliation{IBM Research -- Zurich, S\"{a}umerstrasse 4, 8803 R\"{u}schlikon, Switzerland.}\affiliation{Department of Information Technology and Electrical Engineering, ETH Z\"{u}rich, Gloriastrasse 35, 8092 Z\"{u}rich, Switzerland.}

\author{Michael Hersche} \affiliation{IBM Research -- Zurich, S\"{a}umerstrasse 4, 8803 R\"{u}schlikon, Switzerland.}\affiliation{Department of Information Technology and Electrical Engineering, ETH Z\"{u}rich, Gloriastrasse 35, 8092 Z\"{u}rich, Switzerland.}

\author{Luca Benini} \affiliation{Department of Information Technology and Electrical Engineering, ETH Z\"{u}rich, Gloriastrasse 35, 8092 Z\"{u}rich, Switzerland.}

\author{Abu Sebastian} \email{ase@zurich.ibm.com}\affiliation{IBM Research -- Zurich, S\"{a}umerstrasse 4, 8803 R\"{u}schlikon, Switzerland.}

\author{Abbas Rahimi} \email{abr@zurich.ibm.com} \affiliation{IBM Research -- Zurich, S\"{a}umerstrasse 4, 8803 R\"{u}schlikon, Switzerland.}

\date{\today}

\begin{abstract}
Disentanglement of constituent attributes of a sensory signal is central to sensory perception and cognition and hence is a critical task for future artificial intelligence systems. In this paper, we present a compute engine capable of efficiently factorizing high-dimensional holographic representations of combinations of such attributes by exploiting the computation-in-superposition capability of brain-inspired hyperdimensional computing and the intrinsic stochasticity associated with analog in-memory computing based on nanoscale memristive devices. Such an iterative in-memory factorizer is shown to solve at least five orders of magnitude larger problems that cannot be solved otherwise, while also significantly lowering the computational time and space complexity. We present a large-scale experimental demonstration of the factorizer by employing two in-memory compute chips based on phase-change memristive devices. The dominant matrix-vector multiply operations take a constant time irrespective of the size of the matrix ($\bigO(1)$) thus reducing the computational time complexity to merely the number of iterations. Moreover, we experimentally demonstrate the ability to factorize visual perceptual representations reliably and efficiently.
\end{abstract}
\maketitle
\section*{Introduction}

One of the fundamental problems in sensory perception is unbinding---the separation of causes of a raw sensory signal~\cite{Feldman2013} that contain multiple attributes. For instance, the pixel intensities sensed by photoreceptors result from the combination of different physical attributes ~\cite{Land1971,Tenenbaum1978,Adelson1996,Feldman2013,Barron2015}. For example, the observed luminance at a point on the sensor is a multiplicative combination of reflectance and shading~\cite{Tenenbaum1978}. To be able to estimate these constituent factors, visual perception must begin with the observed luminance and solve an inverse problem that involves undoing the multiplication by which the attributes were combined~\cite{Adelson1996,Feldman2013}. This factorization problem is also at the core of other levels of the conceptual hierarchy, such as factoring time-varying pixel data of dynamic scenes into persistent and dynamic components~\cite{Hinton2010,Sompolinsky2010,Olshausen2012,Olshausen2020}, factoring a sentence structure into roles and fillers~\cite{SmolenskyTensor1990,Jackendoff2002}, and finally cognitive analogical reasoning~\cite{Hummel1997,KanervaAnalogy1998,KanervaPattern1998,PlateAnalogy2000,GaylerIsomorphism2009}. How these factorization problems could be solved efficiently by biological neural circuits is still unclear to date. Moreover,  given their ubiquitous presence in perception and cognition, it is essential that future artificial intelligence systems are equipped with compute units that can perform these factorization operations efficiently across very large problem sizes.

An elegant mathematical approach to representing the combination of attributes is via high-dimensional holographic vectors in the context of brain-inspired vector symbolic architectures~\cite{GaylerJackendoff2003,PlateHolographic1995,PlateHolographic2003,KanervaHyperdimensional2009} (see Supplementary Note~1).
They are holographic since the encoded information is distributed equally over all the components of the vector. Moreover, any two randomly drawn vectors, by virtue of their high dimensionality, are almost orthogonal to each other, i.e., their expected similarity is close to zero with a high probability~\cite{KanervaHyperdimensional2009}. These vectors can also be manipulated by a rich set of dimensionality-preserving algebraic operations. In one approach, an object with $F$ attributes can be described by element-wise multiplication of an associated $D$-dimensional holographic bipolar ($\{-1, +1\}^{D}$) vector corresponding to each attribute, which results in a unique product vector of the same fixed dimensionality~\cite{FradyResonator2020}. The element-wise multiplication operation can be viewed as the binding operation that binds the attribute vectors and generates the product vector. Moreover, it has recently been shown that, given a raw image of an object, a deep convolutional neural network can be trained to generate the product vector approximately~\cite{NVSA}. The factorization problem can now be posed as the decomposition of an exact product vector or, as in the latter case, an inexact product vector, into its constituent attribute vectors.

\begin{figure}[t]
\includegraphics[trim={0 6.20cm 0 0},clip,width=\textwidth]{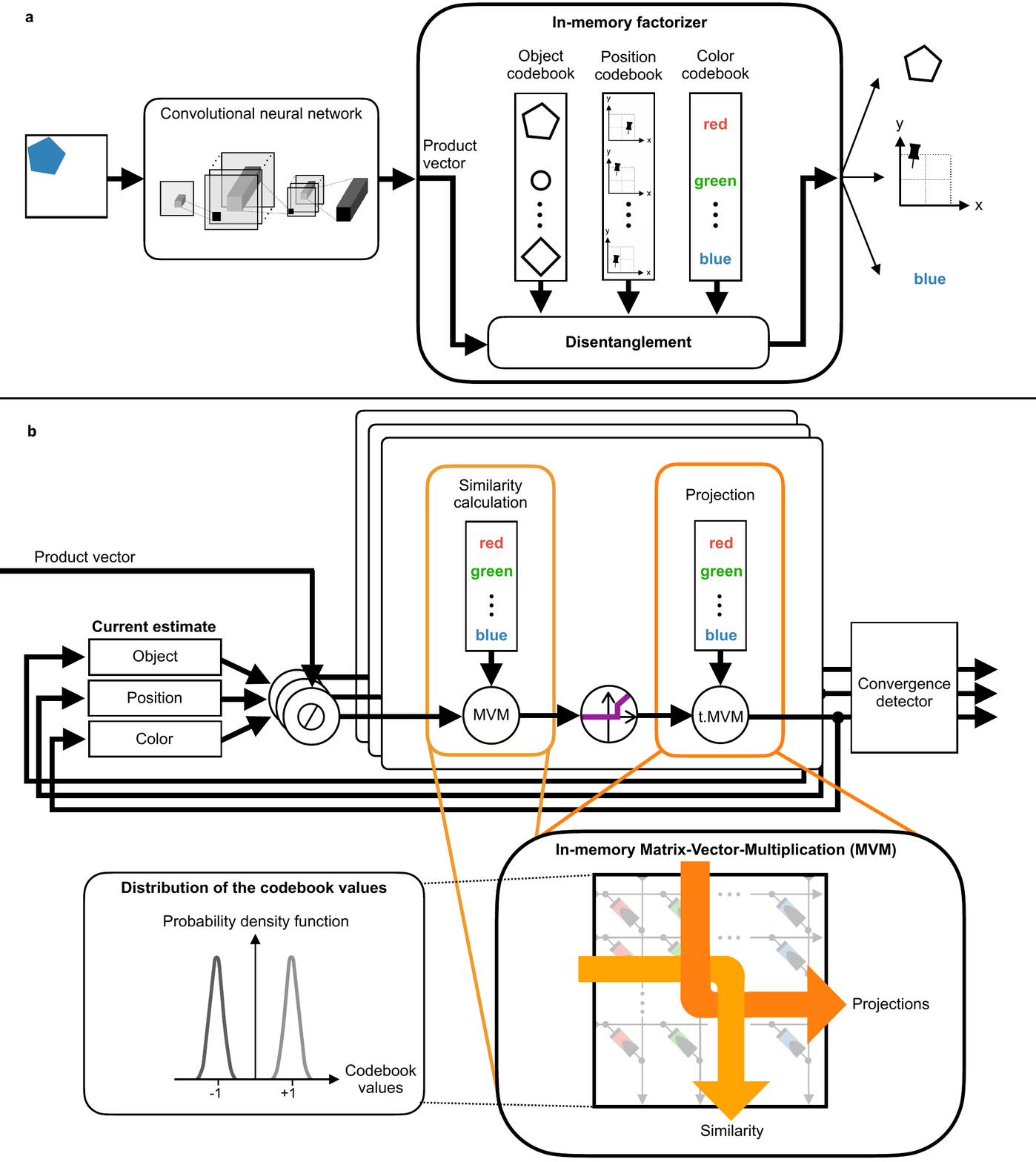}
\caption{
\textbf{Factorization of perceptual representations using the in-memory factorizer.}
(\textbf{a}) The visual input is first fed through a convolutional neural network to approximately map it to a $D$-dimensional product vector. The generated product vector is applied as an input to the in-memory factorizer, which contains a unique codebook of codevectors for each possible attribute. The factorizer disentangles the product vector and predicts the correct attribute factors. (\textbf{b}) The in-memory factorizer iteratively searches in superposition. For each attribute, an updated estimate is computed during every iteration by unbinding the contribution of the other factors from the product vector. The unbound estimate is fed through the similarity calculation with the sparse activation as nonlinearity, and the projection to obtain a novel estimate, which is then fed back to be used in the subsequent iteration. Similarity calculation is based on MVM operations, and projection is based on transposed MVM operations that can both be executed in-memory in a crossbar array of memristive devices by exploiting the Ohm's law and Kirchhoff's current summations law with a computational time complexity of $\bigO(1)$. Moreover, the intrinsic stochasticity associated with storing the bipolar codevectors in the array and the resulting imprecise MVM operations serves as a key enabler for the in-memory factorizer.}
\label{fig:fig1}
\end{figure}

In this article, we propose a non-deterministic, non-von Neumann compute engine that efficiently factorizes such product vectors to obtain estimates of the constituent attributes. The compute engine combines the emerging compute paradigm of in-memory computing (IMC)~\cite{MemristiveTech_Science2022,IMC_NatureNano2020,IMC_NatRevMat2020} with an enhanced variant of a resonator network~\cite{FradyResonator2020,KentResonatorNetworks2020}. 
The resonator network is a nonlinear dynamical system capable of factorizing holographic vectors and is indeed a viable neural solution to the factorization problem~\cite{FradyResonator2020}. The resonator network searches for the $F$ attributes across the set of possible solutions referred to as the codebook associated with each attribute. The vectors associated with each codebook are referred to as codevectors (see Fig.~\ref{fig:fig1}a, with $F=3$). All codevectors are randomly drawn which makes them quasi-orthogonal to each other in the high-dimensional space as was mentioned earlier~\cite{KanervaHyperdimensional2009}. When each codebook contains a finite set of $M$ codevectors, there are $M^F$ possible combinations to be searched in order to factorize the $D$-dimensional product vector into its constituent factors, where $D << M^F$. Factorizing the product vectors constructed by binding randomly drawn vectors that exhibit no correlational structure forms a hard combinatorial search problem. By exploiting the quasi-orthogonality of the codevectors, the resonator network is able to rapidly search these many combinations in superposition by iteratively unbinding all but one of the factors with the product vector, and then projecting it into the space of possible solutions of the considered factor. Note that, in the bipolar space, unbinding is also performed via element-wise multiplication. Both the similarity search and projection operations associated with the resonator network involve matrix-vector multiply (MVM) operations where the matrix transpires to be a fixed codebook. This is highly amenable to IMC using memristive devices~\cite{Wong2015,Chua2011}. Moreover, the intrinsic stochasticity of the devices could be a valuable computational tool as has been proposed for solving energy-based combinatorial optimization problems such as simulated annealing~\cite{SimAnnealing_IEDM2018}, Boltzmann machines~\cite{RBM_HPCA2016,Boltzman_NatCom_2019,Ising_Nature2019,RBM_ISSCC2020}, and Hopfield networks~\cite{Hopfield_Nature2017,Hopfield_NatElec2020,Hopfield_SicAdv2020}. 

The proposed in-memory factorizer stores the codevectors on crossbar arrays of memristive devices performing analog in-memory MVM operations. As shown in Fig.~\ref{fig:fig1}b, the similarity calculation and projection are based on MVM and transposed MVM operations, respectively. These operations can be executed in-memory in a crossbar array of memristive devices by exploiting the Ohm’s law and Kirchhoff’s current summations law. Moreover, the use of a nonlinear function (as sparse activations) between MVM and transposed MVM operations, and the intrinsic stochasticity associated with the memristive devices can enhance the maximally solvable problem size. Finally, we present a large-scale experimental demonstration using in-memory compute cores based on phase-change memory (PCM) technology and applications in visual perception.

\section{In-memory stochastic factorizer with sparse activations}
\begin{figure}[t]
\includegraphics[trim={0 23cm 0 0},clip,width=\textwidth]{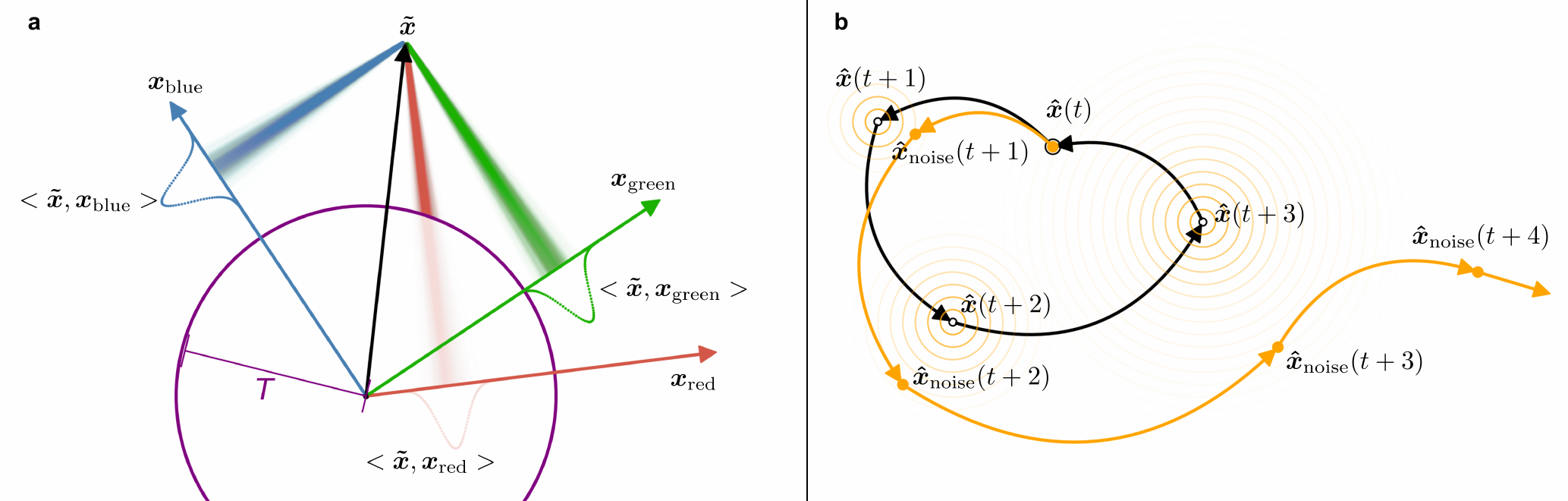}
\caption{\textbf{Stochastic similarity computation, sparse activations, and limit cycles.}
(\textbf{a}) The similarity calculation computes the similarities between the unbound estimate vector $\boldsymbol{\Tilde{x}}$ and all the codevectors (e.g., $\{\boldsymbol{x}_{\mathrm{red}}, \boldsymbol{x}_{\mathrm{green}}, \boldsymbol{x}_{\mathrm{blue}}\}$). The in-memory similarity calculation, denoted by $<.,.>$, is stochastic with additive noise. The distribution of noisy similarity results projected onto the 2-D space is shown in green, red, and blue colors respectively. Further, a winners-take-all approach \emph{activates} similarity values only above a certain threshold ($T$) level which results in a similarity vector with sparse non-zero elements.
Similarity values smaller than $T$ are zeroed out, for instance, $<\boldsymbol{\Tilde{x}},\boldsymbol{x}_{\mathrm{red}}>$ is not activated (i.e., zeroed), while the other two similarity values ($<\boldsymbol{\Tilde{x}},\boldsymbol{x}_{\mathrm{blue}}>$ and $<\boldsymbol{\Tilde{x}},\boldsymbol{x}_{\mathrm{green}}>$), larger than $T$, remain activated. The activation threshold ($T$) is depicted in purple. (\textbf{b}) Visualized by the black arrows, this figure shows a limit cycle of length $l$=$4$. When stuck in a limit cycle of length $l$, we end up constantly checking the same $l$ solutions. In contrast, the orange arrows show an example trajectory of the factor's estimates ($\boldsymbol{\hat{x}}$) of the in-memory factorizer exploiting stochasticity in both similarity and projection operations which yields noisy estimates. For a subsequent time step, there is some uncertainty as visualized by the orange circles. As the search for factorization is an iterative process, the uncertainty for the subsequent time steps increases. Eventually, the uncertainty is high enough for the in-memory factorizer to diverge from a limit cycle, and to converge to the correct factorization in time.}
\label{fig:fig2}
\end{figure}

The unsupervised nature of the conventional resonator network's deterministic search procedure could result in checking the same sequence of solutions multiple times across iterations, resulting in limit cycles that prevent convergence to the optimal solution. One of the key insights from the in-memory factorizer is that the intrinsic stochasticity associated with the memristive devices can significantly reduce the occurrence of such limit cycles. As shown in Fig.~\ref{fig:fig2}a, during the similarity calculation, the analog in-memory MVM results in a stochastic similarity vector. The stochasticity enables the factorizer to break free of limit-cycles and thus explore a substantially larger solution space (Fig.~\ref{fig:fig2}b). 

Another limitation of the resonator network is the use of the identity function, as a linear activation, between the similarity search and the projection operations. We find that adopting a nonlinear winners-take-all approach, by zeroing out the weaker similarity values, enhances both the convergence rate and maximally solvable problem size of the in-memory factorizer (see Supplementary Note~2). The winners-take-all approach uses an activation threshold, $T$, to sparsify the similarity vector (Fig.~\ref{fig:fig2}a) which was chosen based on Bayesian optimization (see Methods). 

\begin{figure}[t]
\includegraphics[trim={0 15.5cm 0 0},clip,width=\textwidth]{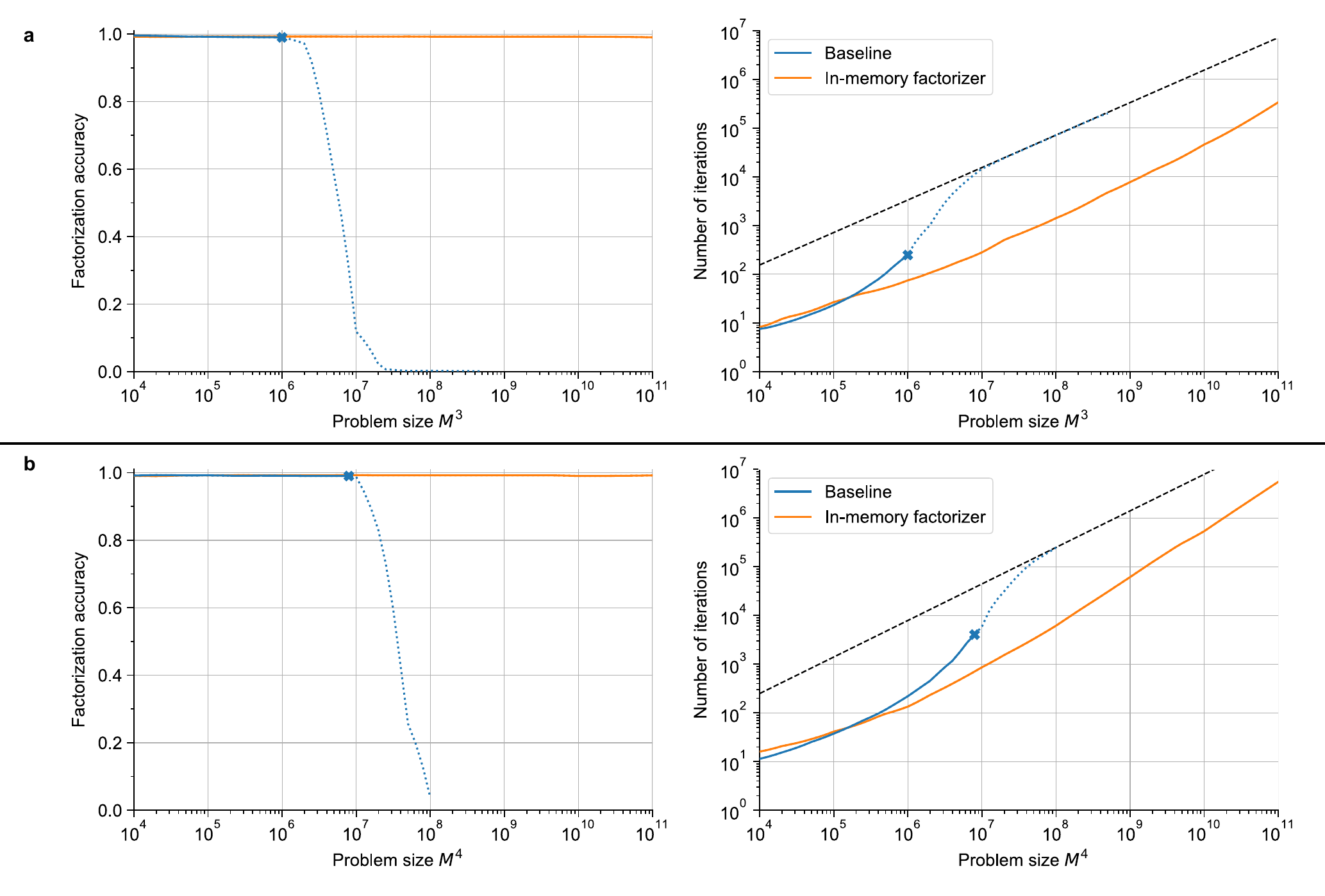}
\caption{
\textbf{Operational capacity of the stochastic in-memory factorizer with sparse activations.} This figure shows the operational capacity for: $F$=$3$ in (\textbf{a}), and $F$=$4$ in (\textbf{b}). The problem size $M^F$ is shown on the X-axis, and the Y-axis shows the accuracy on the left side of the plot, and the number of iterations required to solve a given problem size on the right side. The black dashed lines indicate the equivalent number of dot-product operations required for a brute-force approach to search among $M^F$ precomputed product vectors. These lines indicate the upper limit for the operation count. For each $F$, we use the smallest dimension reported by the baseline resonator network~\cite{FradyResonator2020,KentResonatorNetworks2020}, namely $D$=$1500$ for $F$=$3$, and $D$=$2000$ for $F$=$4$. The blue cross indicates the largest problem size that is within the operational capacity of the baseline resonator network, meaning that the problems larger than that size can not be factorized by the baseline network at 99\% accuracy, while the in-memory factorizer can solve at least five orders of magnitude larger problem sizes at 99\% accuracy, or higher.}
\label{fig:fig3}
\end{figure}

The resulting non-deterministic in-memory factorizer with nonlinear sparse activations can be analyzed and compared with the state-of-the-art using three figures of merit: the dimensionality, the computational complexity, and the operational capacity. The dimensionality refers to the number of elements in a codevector. The computational complexity defines the average number of the operations required by a factorizer to factorize a given product vector where each operation refers to a $D$-dimensional dot product calculation. It depends on the problem size and the number of factors. Given an upper bound for the first two figures of merit, the operational capacity defines the maximally solvable problem size with an accuracy of at least 99\%. To compare the operational capacity of the in-memory factorizer with the baseline resonator network~\cite{FradyResonator2020,KentResonatorNetworks2020} we performed extensive simulations of the in-memory factorizer in software (see Methods). As seen in Fig.~\ref{fig:fig3}, the in-memory factorizer significantly enhances the operational capacity by up to five orders of magnitude even when the vector dimensionality is reduced by a factor of over four. (Fig.~S1). 

\section{Large-scale experimental demonstration}
\begin{figure}[t]
\includegraphics[trim={0 5.2cm 0 0},clip,width=0.9\textwidth]{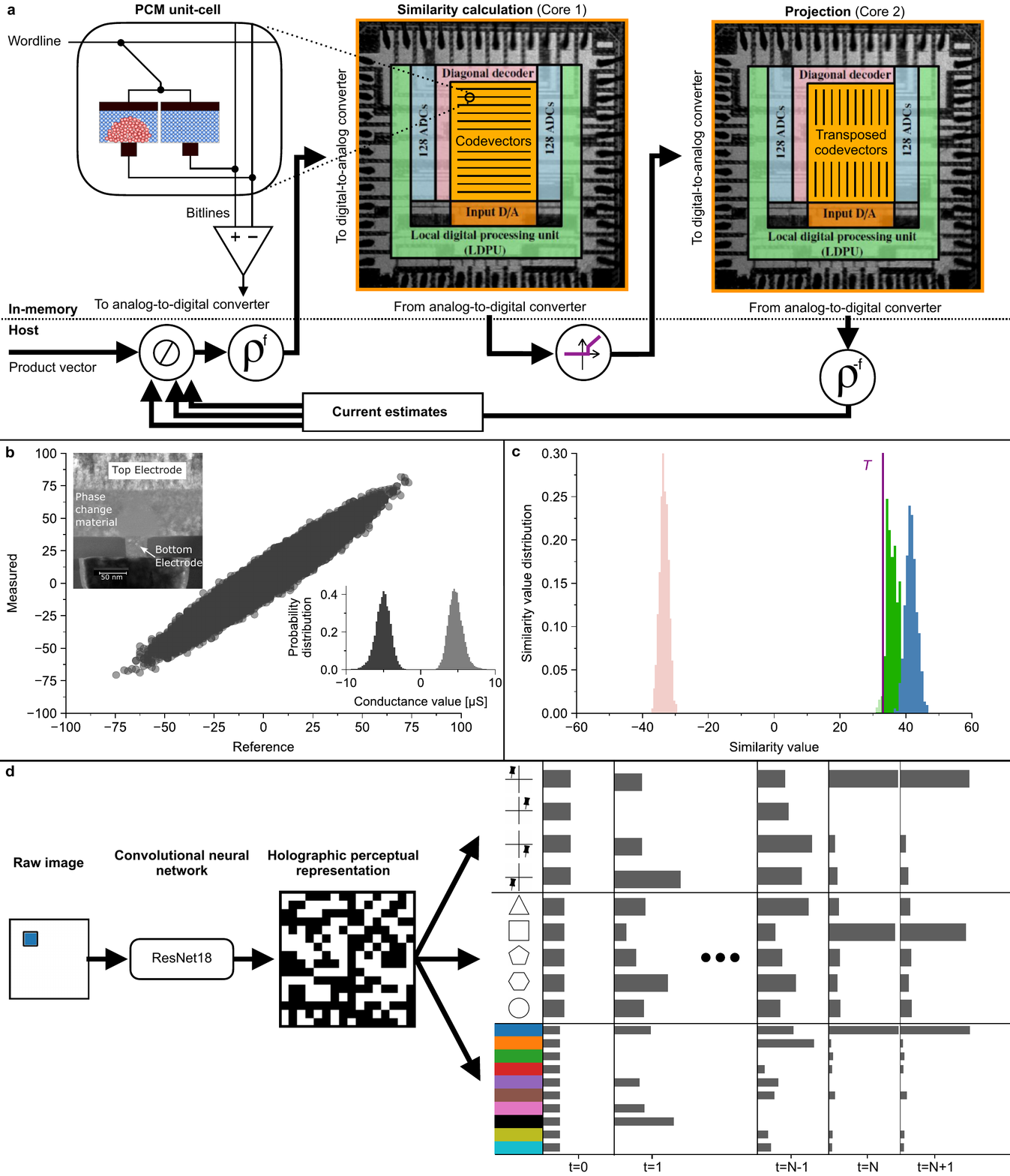}
\caption{
\textbf{Experimental realization of the in-memory factorizer.} (\textbf{a}) The experimental setup includes two in-memory compute cores for similarity calculation and projection. Each MVM operation is executed using the crossbars and the fully integrated peripherals on one of the two cores. The unbinding, permutation, activation, and bipolarization operations are executed on the host computer connected to the setup. On the crossbar, one unit cell stores one bipolar ($\pm1$) weight value associated with the codevectors. In the first core, the codevectors are stored along the columns of the crossbar, while in the second core, the codevectors are stored along the rows. Each unit-cell is programmed to the desired value. (\textbf{b}) The measured output of the in-memory MVM plotted against the reference output obtained in high precision indicating the intrinsic stochasticity associated with the operation. The gray histogram corresponds to the positive conductance values, and the black histogram to the negative ones. The bottom-right inner panel shows the distribution of the programmed conductance values on the first core executing the similarity calculation. The top-left inner panel shows a Low-angle annular darkfield (LAADF) scanning transmission electron microscope (STEM) image \cite{Y2021liISTFA} of a PCM device in its RESET state where a large amorphous region blocks the bottom electrode resulting in a conductance value close to zero. To encode a 1 or -1 in a unit-cell, the amorphous region corresponding to the appropriate device is partially crystallized to achieve a higher conductance value. The target conductance value was set at $5\mu$S. (\textbf{c}) The red, green, and blue histograms correspond to the distribution of similarity between the same unbound estimate vector ($\Tilde{x}$) and three color codevectors of $x_{\mathrm{red}}$, $x_{\mathrm{green}}$, $x_{\mathrm{blue}}$, respectively. The purple vertical line corresponds to the activation threshold at $T$=$33$, which avoids activating any similarity values related to $x_{\mathrm{red}}$. (\textbf{d}) The raw image is fed through a ResNet18 to generate a holographic product vector visualized as a binary heatmap. The in-memory factorizer iteratively disentangles the holographic vector. In the first time step, all codevectors show an equal similarity to the current estimate. Over time, the in-memory factorizer converges to the correct factorization, indicated by a high similarity value for a single codevector.}
\label{fig:fig4}
\end{figure}

Next, we present an experimental realization of the in-memory factorizer using IMC cores based on PCM devices fabricated in \unit[14]{nm} CMOS technology (see Methods). We employed two IMC cores, one to calculate the similarities and one for the projections (Fig.~\ref{fig:fig4}a). Each IMC core features a crossbar array of $256\times256$ unit-cells capable of performing stochastic and analog MVM operations~\cite{9508706}. Each unit-cell comprises four PCM devices organized in a differential configuration and can be programmed to a certain conductance value (the bipolar codebooks are stored in two out of four devices). However, due to the intrinsic stochasticity associated with crystal nucleation and growth \cite{Y2016tumaNatureNano}, there will be a distribution of conduction values across multiple devices in the crossbar (Fig.~\ref{fig:fig4}b). This distribution will get slightly broader with time as a result of the variability associated with the structural relaxation of the atomic configuration in each device~\cite{Y2018legalloAEM}. In addition to this, there is read noise that exhibits a $1/f$ spectral characteristic and random telegraph noise (see Supplementary Note~3 for more details)~\cite{Y2020legalloJPD}. The input to the IMC core is applied using a constant pulse-width modulated voltage applied to all the rows of the crossbar array in parallel. Ohm's law defines the current flowing through each unit-cell and the current is summed up on the corresponding bitlines in accordance with Kirchhoff's current summation law. This current is digitized and accumulated by 256 analog-to-digital converters operating in parallel. 

Each IMC core performs MVM operations in a constant amount of time, $\bigO(1)$, which leads to reducing the time complexity of factorization to merely the average number of iterations. Moreover, the intrinsic randomness associated with the PCM devices is exploited to calculate the stochastic similarity and projection vectors that minimizes the occurrence of the limit cycles  (Fig.~\ref{fig:fig4}b and c). A permute logic is employed to temporally multiplex one crossbar array for multiple factors and the hyperparameters such as the activation and convergence thresholds were obtained through Bayesian learning (see Methods). The experimentally realized in-memory factorizer is compared to the baseline resonator network~\cite{FradyResonator2020,KentResonatorNetworks2020}, whereby both methods used $D=256$, $M=256$, and $F=3$. These parameters set the total problem size to $M^F=16,777,216$, and the maximum number of iterations to $N=21,845$, which ensures that the computational complexity does not exceed that of the brute-force search (see Methods). When using the same codebooks and $5,000$ randomly selected product vectors as queries, the baseline resonator network is found to be incapable of factorizing any of the product vectors. In contrast, the in-memory factorizer is capable of achieving an outstanding accuracy of 99.71\% with an average number of $3,312$ iterations. 

Finally, we demonstrate the role of the in-memory factorizer in visual perception to disentangle the attributes of raw images. The perception system consists of two main components, as shown in Fig.~\ref{fig:fig4}d. A convolutional neural network is trained to map the input image to a holographic perceptual product vector, based on a known set of image attributes. Hence, during inference, the output of the neural network is an approximation of the product vector that describes the image. The in-memory factorizer is used to disentangle the approximate product vector using the known set of image attributes.
For experiments, we used the input images from the relational and analogical visual reasoning (RAVEN) dataset~\cite{zhang2019raven}. The images contain objects with attributes such as type, size, color, and position. Each set of attributes is mapped to a unique codebook leveraging its symbolic nature (see Methods). The disentanglement of $1,000$ images from the RAVEN dataset to the correct estimate of the attributes achieved an accuracy of 99.42\%.
\section{Discussion}
We compare the in-memory factorizer with a dedicated reference digital design that benefits from the proposed sparse activations and all other features, except the intrinsic stochasticity that is inherent to the PCM devices. Using the same configurations of the experiments ($D=256$, $M=256$, $F=3$, and $5,000$ trials), the deterministic digital design was able to reach an accuracy of 95.76\% with $3,802$ iterations on average. Compared to the in-memory factorizer, the digital design demanded 14.8\% more iterations on average, and still exhibited 4\% lower accuracy. Even if we allow the maximum number of iterations to be arbitrarily large, the digital design is still not able to match the accuracy of the in-memory factorizer due to the limit cycles (see Supplementary Note~4) thus highlighting the crucial role of the intrinsic stochasticity associated with the PCM devices.

The in-memory factorizer could also result in a significant gain in energy and areal efficiency. By combining the advantages of in-place computation and reduced number of iterations, it is estimated that a custom-designed in-memory factorizer based on a $512\times 512$ crossbar array is capable of factorizing a single query within an energy budget of 33.1\,$\mu$J on average, resulting in energy savings of 12.2\,$\times$ compared to the reference digital design. The total area saving is estimated to be 4.85$\times$ (see Supplementary Note~4). The non-volatile storage of the codebooks is an added advantage compared to the digital design.

Note that the application of in-memory factorizers can go beyond visual perception, as factorization problems arise everywhere in perception and cognition, for instance in analogical reasoning~\cite{Hummel1997,KanervaAnalogy1998,KanervaPattern1998,PlateAnalogy2000,GaylerIsomorphism2009,SpencerPhDThesis}. Other applications include tree search~\cite{FradyResonator2020} and the prime factorization of integers~\cite{IntegerFacRes_NICE2022}. Recently, it has been shown how the classical integer factorization problem can be solved by casting it as a problem of factorizing holographic vectors~\cite{IntegerFacRes_NICE2022}. These pave the way for solving non-trivial combinatorial search problems.

To summarize, we have presented a non-von Neumann compute engine to factorize high-dimensional, holographic vectors by searching in superposition. We have experimentally verified it using two state-of-the-art in-memory compute cores based on PCM technology which provides in-place computation and non-volatility. The experimental results show the reliable and efficient factorization of holographic vectors spanning a search space of $256\times256\times256$. The intrinsic stochasticity associated with the PCM devices is coupled with the nonlinear sparse activations to enhance the operational capacity by five orders of magnitude as well as to reduce the spatial and time complexity associated with the computational task. Furthermore, we demonstrated the efficacy of the in-memory factorizer in disentangling the constituent attributes of raw images. This work highlights the role of emerging non-von Neumann compute paradigms in realizing critical building blocks of future artificial intelligence systems. 

\section*{Methods}
\subsection*{{Detection of convergence in the in-memory factorizer}}
The iterative factorization problem is said to be converged if, for two consecutive time steps, all the estimates are constant, i.e., $\hat{\boldsymbol{x}}^{\text{f}}(t+1) = \hat{\boldsymbol{x}}^{\text{f}}(t)$ for $f$ $\in$ $[1,F]$.  To detect this convergence, we define a novel early convergence detection algorithm. The in-memory factorizer is said to be converged if a single similarity value across all the factors surpasses a convergence detection threshold: 
\begin{equation}\label{eq:convergence}
\text{converged} =\begin{cases}
    \text{true}, & \text{if $\alpha^{\text{f}}_{\text{i}}>T_{\text{convergence}}$}\\ 
    \text{false}, & \text{otherwise},
  \end{cases}
\end{equation}
where $i \in [1,|X^{\text{f}}|]$ for $f$ $\in$ $[1,F]$. 
Upon convergence, the predicted factorization is given by the codevector associated with the largest similarity value per codebook. The novel convergence detection counteracts parasitic codevectors with a high enough similarity value to prevent the in-memory factorizer from converging to a stable solution.
Furthermore, there is no need to store the history of prior estimates.
Otherwise, the last estimate for each factor had to be stored to be able to compare it to the latest estimate and detect convergence, resulting in a total of $F\cdot D$ stored bits. We used Bayesian optimization to find the optimal convergence detection threshold. It stays at a fixed ratio of $D$ for any given set of hyperparameters and problem sizes.

\subsection*{{Maximal number of iterations}}
To use strictly less search operations compared to the brute-force approach, the maximal number of iterations ($N$) must be constrained to: 
\begin{equation}\label{eq:maxItr}
    N \cdot M \cdot F < M^{\text{F}}.
\end{equation}
In all our experiments, we constrain the maximum number of iterations to $N < \frac{M^{\text{F}-1}}{F}$ to ensure a lower computational complexity compared to the brute-force approach.

\subsection*{{Hyperparameter optimization via Bayesian optimization}}
We find the optimal hyperparameters of the in-memory factorizer $\boldsymbol{h}^{\star} = [T^{\star},T^{\star}_{\text{convergence}}]$ by solving a Bayesian optimization problem where $T^{\star}$ is the activation threshold and $T^{\star}_{\text{convergence}}$ is the convergence threshold. We define the error rate $l$ to be the ratio of the wrongly factorized product vector. More formally, we try to find the optimal set of hyperparameters $\boldsymbol{h}^{\star}$ by minimizing the error rate $l$, which is a function of the hyperparameters $\boldsymbol{h}$: 
\begin{align}
    \boldsymbol{h}^{\star}= arg\,\min_{\boldsymbol{h}}\,l(\boldsymbol{h}).
\end{align}
We model the error rate as a Gaussian process with a radial basis function kernel and minimize it using Bayesian optimization. To sample possible hyperparameters during the optimization we use the expected improvement acquisition function.

We reduce the computational complexity of the error rate evaluation for a given set of hyperparameters $\boldsymbol{h}$ by limiting the maximum number of iterations to $N^{'} = N/10$ and the number of trials to 256. A low number of iterations $N'$ yields higher error rates, yet they still provide good indications for the in-memory factorizer's performance given a set of hyperparameters. Additionally, the reduced number of trials gives noisy evaluations of $l$, which is modeled as additive Gaussian measurement noise. 

We derived the final parameter estimates by averaging over the best five experiments. For the hardware experiments, the stochasticity is inherently provided by the PCM crossbar arrays. However, for simulating the in-memory factorizer in software, we need to model stochasticity as a noise level. Hence, the noise level ($n^{\star}$) is treated as an extra hyperparameter for optimization. Accordingly, for simulating our method in the software we optimized for three hyperparameters: the activation threshold, the convergence threshold, and the noise level given by
$\boldsymbol{h}^{\star} = [T^{\star},T^{\star}_{\text{convergence}},n^{\star}]$.

\subsection*{{In-memory experiments}}
For the experimental demonstration, we employed an in-memory compute core fabricated by IBM Research in \unit[14]{nm} CMOS technology node~\cite{9508706}. It features $256\times 256$ unit cells each comprising four PCM devices arranged in a differential configuration where one pair of devices is connected in parallel to represent positive conductance and another pair to represent negative conductance on the unit cell. For these experiments, however, we program only one device, either on the positive or negative side, as it provides a sufficient dynamic range in conductance to achieve more than 99\% accuracy. This effectively allows us to optimize the unit cell to consist of just two PCM devices, as shown in Fig.~\ref{fig:fig4}a.

A custom-printed circuit board houses two such HERMES cores and an FPGA board is used to control the communication protocol and load data to and from the cores. The FPGA in turn is controlled by a Linux machine running a Python environment on top. The host machine performs the unbinding and applies the activation function, while the two cores perform the dominant MVM similarity search and the projection operations, respectively. An iterative programming scheme of the PCM devices is employed to store the codevectors in the crossbars. The output of the in-memory MVM is measured in terms of the ADC count units. Subsequently, a linear correction is applied to correct circuit-level mismatches. The linear correction parameters are calculated prior to MVM operations. 

The single crossbar core with the dimension of $256 \times 256$ limits the total number of the supported codevectors across all the codebooks to $256$. To overcome this limitation, we propose a permute logic to temporally multiplex one single crossbar array for all $F$ factors. This enables us to exploit the complete crossbar for one single codebook with up to $256$ codevectors, and reuse it across an arbitrary number of factors. As shown in Fig.~4a, prior to the similarity calculation, we apply the permute logic as a factor-wise exclusive circular shift on the estimated unbinding. This results in a quasi-orthogonal time multiplexing of the crossbar. Before updating the estimates, we reverse the circular shift to obtain unaltered estimates.

\subsection*{{Software simulations of the in-memory factorizer}}
The stochasticity is the key enabler of the in-memory factorizer. Adding some stochasticity helps to diverge from the limit cycles as each solution becomes unique. For the experiments reported in Fig.~3, this important aspect is modeled in software by simulating the noisy behaviour of the MVMs on the crossbar as an additive Gaussian noise with zero mean:
\begin{equation}\label{eq:noise}
\alpha_{\text{i}}^{'} = f(\alpha_{\text{i}}) = \alpha_{\text{i}} + n,
\end{equation}
where $\alpha_{i}$ is a single entry of the output vector, and $n$ is normally distributed with $n \sim \mathcal{N}(0,\,\sigma^{2})$.

In total, we simulated $F\times (M+D)$ additive Gaussian noise sources: $M$ on the similarity vector and $D$ on the projection vector. The noisy similarity vector is required to break free of the limit cycles. Due to the random distribution of the similarity values, there is always a chance of activating none of the similarity values if they do not cross the activation threshold. Adding noise on top of the projection prevents such an all-zero estimation by randomly initializing the vector prior to the bipolarization. 

The PCM devices on the crossbar array exhibit similar noisy behavior, which can be modeled as a combination of noise components such as programming noise, read noise and drift variability. In Supplementary Note~3, we model the extent to which each of these noise components is present in the experimental crossbar arrays and analyze the sensitivity to change in the noise components as reflected in the performance figures such as factorization accuracy and the number of iterations required for convergence.

To concretely quantify the effect of the aggregated PCM device noise on the performance of the factorizer, we conducted simulations where the aggregated noise standard deviation gradually increases from zero noise while maintaining the ratio of the standard deviation between read noise to programming noise as observed on the experimental platform ($\sigma_\text{r}/\sigma_\text{p}$=0.3951/1.1636). These results are shown in Extended Data Fig.~\ref{fig:extdata}.

We observe that with zero noise, the factorizer performs poorly with an accuracy of 25.4\% while requiring on average 16,000 iterations to converge. This expected behavior is due to the deterministic nature of the search and the resulting inability to break free of the limit cycles. The factorizer, however, operates at its peak performance when the standard deviation of the aggregated noise is maintained within the range [0.293~$\mu$S, 1.277~$\mu$S]. Note that the standard deviation of the aggregated noise observed on the experimental platform (0.98~$\mu$S) falls in the middle of this tolerated noise range (see Extended Data Fig.~\ref{fig:extdata}). 

\subsection*{{Visual disentanglement}}
We use the RAVEN~\cite{zhang2019raven} dataset, which provides a rich set of Raven’s progressive matrices tests, for visual abstract reasoning tasks. We only focus on the visual perception part of the task to disentangle an image sensory input from its underlying attribute factors. The RAVEN dataset provides a total of 70,000 tests, each consisting of 16 panels of images. In our experiment, we considered the 2x2 image constellation with a single object. Each object can have one of 4 possible positions, 10 colors, 6 sizes, and 5 types. Therefore, there are 1,200 possible combinations. We mapped each set of attributes to a single codebook of the in-memory factorizer. Here, we reused the same codebook as for the synthetic experiment. The first codebook represents the position attribute, the second one represents the color attribute. For the third codebook, we fused the size and type attributes into a single codebook, ending up with a total of 30 possible size-type combinations.

Each image can be described by a product vector formed by the binding of the corresponding codevectors. To map the input image to the product vector, we used a convolutional neural network. We used the ResNet-18 convolutional neural network, and mapped its output to our 256-dimensional vector space using the training schema proposed in~\cite{NVSA}. After training, we ran all the test images though the trained ResNet-18 network to get an estimate of the product vector.

Next, we passed all the estimated product vectors through the same experimental test setup used for the synthetic experiments. The main difference lies in the product vector: while for the synthetic one we used the exact product vectors, for the visual perception task the product vectors are the output of the convolutional neural network, which will be an approximate product vector as opposed to being the exact one.


\section*{Acknowledgements}
This work is supported by the IBM Research AI Hardware Center, and by the Swiss National Science foundation (SNF), grant 200800. The authors would like to thank Manuel Le Gallo for technical help, Kevin Brew and Juntao Li for assistance with TEM imaging of PCM devices and Vijay Narayanan, Chid Apte, and Robert Haas for managerial support.

\section*{Data availability} 
The data that support the findings of this study are available at \href{https://doi.org/10.5281/zenodo.7599430}{https://zenodo.org/record/7599430}.

\section*{Code availability}
Our code is available at \href{https://github.com/IBM/in-memory-factorizer}{https://github.com/IBM/in-memory-factorizer}

\section*{Author contributions}
J.L., G.K., M.H., A.S., and A.R. conceived the idea and designed the experiments. J.L. performed experiments and characterization. J.L., A.S., and A.R. wrote the manuscript with input from all authors.
All authors provided critical comments and analyses.

\clearpage
\clearpage
\setcounter{figure}{0}
\renewcommand{\figurename}{Extended Data Fig.}
\begin{figure}[t]
\includegraphics[width=0.5\textwidth]{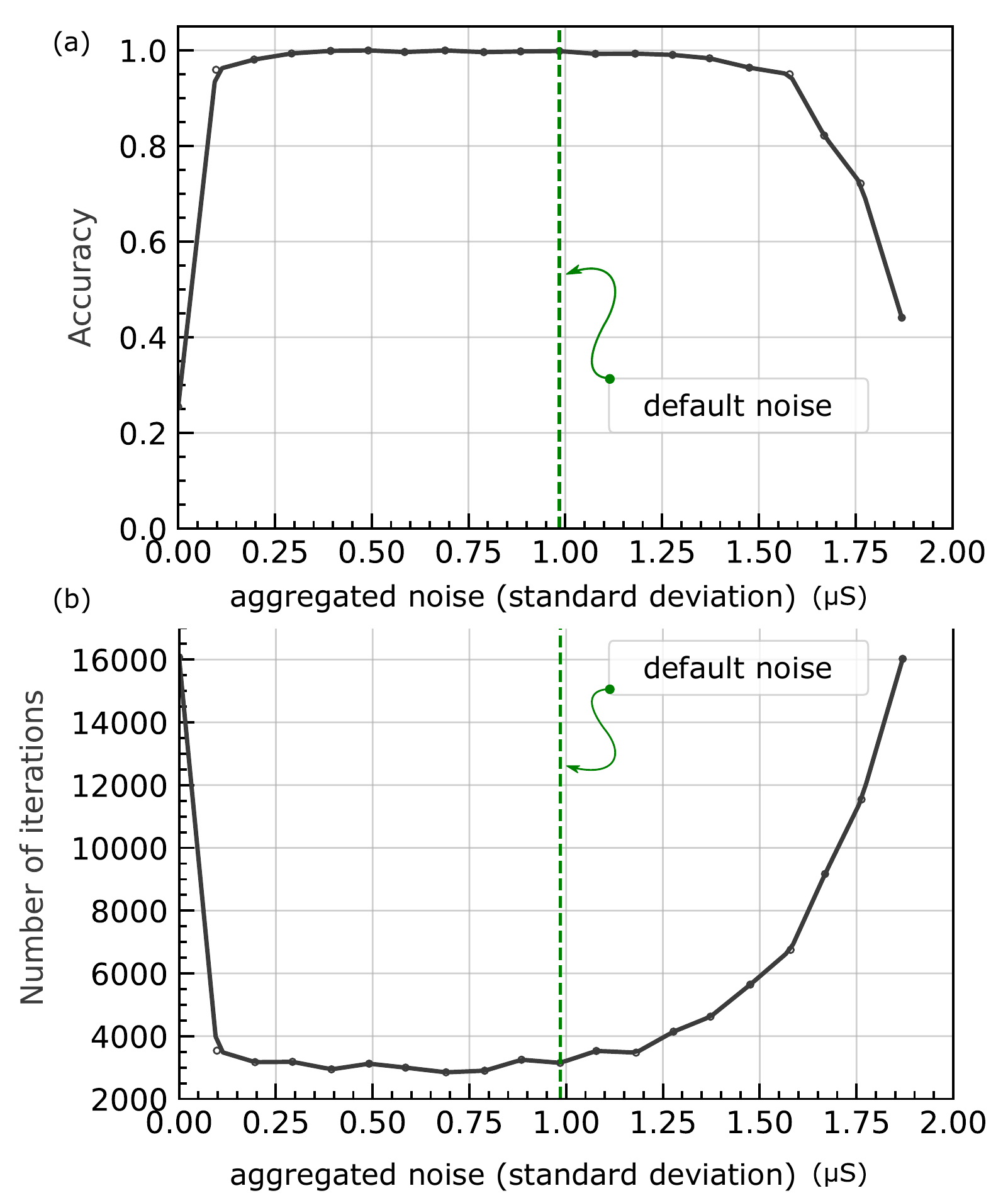}
\caption{\textbf{Desirable range of noise.} The aggregated noise corresponding to the programming noise, drift variability, and read noise in the PCM devices affects (a) the accuracy of factorization, and (b) the number of iterations to converge. The optimal range for the standard deviation of the noise lies between 0.293~$\mu S$ and 1.277~$\mu S$. As indicated by the green vertical line, the level of noise observed in the experimental crossbar array lies within the desirable range of the noise.}
\label{fig:extdata}
\end{figure}

\clearpage

\setcounter{section}{0}
\setcounter{table}{0}
\setcounter{figure}{0}
\setcounter{equation}{0}
\renewcommand{\thesection}{S\arabic{section}}   
\renewcommand{\thetable}{S\arabic{table}}   
\renewcommand{\thefigure}{S\arabic{figure}}
\renewcommand{\theequation}{S\arabic{equation}}
\renewcommand{\figurename}{Fig.~}

\section*{Supplementary Notes}
\subsection*{{Supplementary Note~1: Background in vector symbolic architectures and resonator networks}}
\subsubsection*{Vector symbolic architectures}
Here, we provide a brief overview of vector symbolic architectures (VSAs)~\cite{GaylerJackendoff2003,PlateHolographic1995,PlateHolographic2003,KanervaHyperdimensional2009} of which the resonator networks~\cite{FradyResonator2020,KentResonatorNetworks2020} are based on. VSA is a powerful computing framework that is built on an algebra in which all representations are high-dimensional holographic vectors of the same, fixed dimensionality denoted by $D$. This is attributed to modeling the representation of information in the brain as distributed over many neurons. In this work, we consider a VSA model based on bipolar vector space~\cite{GaylerJackendoff2003}, i.e., $\lbrace -1,+1\rbrace^{D}$. The similarity between two vectors is defined as the cosine similarity:
\begin{equation} \label{eq:similarity}
\text{sim}(\boldsymbol{x}_{1},\boldsymbol{x}_{2}) = \frac{\langle  \boldsymbol{x}_{1} , \boldsymbol{x}_{2} \rangle}{||\boldsymbol{x}_{1}|| ||\boldsymbol{x}_{2}||} = \frac{\langle  \boldsymbol{x}_{1} , \boldsymbol{x}_{2} \rangle}{D}
\end{equation}
As one of the main properties of the high-dimensional vector space, any two randomly drawn vectors lie close to quasi-orthogonality to each other, i.e., their expected similarity is close to zero with a high probability~\cite{KanervaHyperdimensional2009}. The vectors can represent symbols, and can be manipulated by a rich set of dimensionality-preserving algebraic operations:
\begin{itemize}
    \item Binding: Denoted by $\odot$, the Hadamard (i.e., element-wise) product of two input vectors implements the binding operation. It is useful to represent a hierarchical structure whereby the resulting vector lies quasi-orthogonal to all the input vectors. The binding operation follows the commutative law $\boldsymbol{x}_{1} \odot \boldsymbol{x}_{2} = \boldsymbol{x}_{2} \odot \boldsymbol{x}_{1} = \boldsymbol{p}$.    
    \item Unbinding: The unbinding operation reverses the binding operation. As the element-wise multiplication in the bipolar space is self-inverse, the same operation as for the binding can be used. Using the unbinding operator $\oslash$ the operation is defined as $\boldsymbol{p} \oslash \boldsymbol{x}_{1} = \boldsymbol{x}_{2}$.
    
    \item Bundling: The superposition of two vectors is calculated by the bundling operation $\oplus$.
    The operation is defined by an element-wise sum with consecutive bipolarization. In case of an element-wise sum equal to zero, we randomly bipolarize. 
    
    \item Clean-up: The clean-up operation maps a noisy vector to its noise-free representation by an associative memory lookup.
    
    \item Permutation: Permutation is a unary operation on a vector that yields a quasi-orthogonal vector of its input. This operation rotates the coordinates of the vector. A simple way to implement this is as a cyclic shift by one position.
\end{itemize}
For further details, please refer to a detailed survey about VSAs~\cite{HDC_Survey_PI,HDC_Survey_PII}. 

\subsubsection*{Resonator networks for iterative factorization}
Here, we consider a specific form of factorization in which a product vector, based on high-dimensional holographic representations of VSAs, can be factorized into its factors.
For each factor, a set of possible solutions is stored in a so-called codebook.
Each codebook $X$ contains $M$ possible solutions called codevector of $\{-1,+1\}^D$:
\begin{equation}
X := \{\boldsymbol{x}_{1}, \boldsymbol{x}_{2}, \dots, \boldsymbol{x}_{M}\}.
\end{equation} 
The number of factors is defined as $F$.
Multiplicative binding among these codebooks forms a total problem size of $M^F$ to be searched (i.e., all possible combinations of codevectors). The codevectors are drawn randomly (hence they are quasi-orthogonal), and are combined by multiplicative binding that is a \emph{randomizing} operation. This creates a combinatorial search for factorization because the resulting $M^F$ quasi-orthogonal vectors in the search space share no particular correlational structure. A brute-force approach requires searching along all possible combinations.
In contrast, the resonator networks~\cite{FradyResonator2020,KentResonatorNetworks2020} exploit the VSA operations to iteratively search in superposition for factorizing a given product vector. 

Given an entangled (bound) product vector $\boldsymbol{p}$ and the set of codebooks $X^{1}, X^{2}, \dots, X^{\text{F}}$, the resonator network iteratively searches in superposition to find a valid factorization $\boldsymbol{\hat{x}}^{1} \in X^{1}, \boldsymbol{\hat{x}}^{2} \in X^{2}, \dots, \boldsymbol{\hat{x}}^{F} \in X^{F}$ such that the estimated vector $\boldsymbol{\hat{p}} = \boldsymbol{\hat{x}}^{1} \odot \boldsymbol{\hat{x}}^{2} \odot \dots \odot \boldsymbol{\hat{x}}^{\text{f}}$ resembles with the highest similarity to the input product vector $\boldsymbol{p}$. The initial estimates (at $t=0$) for the factors $\hat{\boldsymbol{x}}^{1}(0)$, $\hat{\boldsymbol{x}}^{2}(0)$, \dots, $\hat{\boldsymbol{x}}^{\text{F}}(0)$ are obtained by superposing all the codebook elements. This provides each element in the codebook an equal chance.
At any time step $t$, the unbound estimates are calculated first.  For a given factor, the unbinding is performed by taking the product vector $\boldsymbol{p}$ and unbinding the contribution of the other factors latest estimate:
\begin{equation}\label{eq:res:unbinding}
\begin{split}
\tilde{\boldsymbol{x}}^{1}(t) & = \boldsymbol{p} \oslash \hat{\boldsymbol{x}}^{2}(t) \oslash \hat{\boldsymbol{x}}^{3}(t) \oslash \dots \oslash \hat{\boldsymbol{x}}^{\text{F}}(t) \\
\tilde{\boldsymbol{x}}^{2}(t) & = \boldsymbol{p} \oslash \hat{\boldsymbol{x}}^{1}(t) \oslash \hat{\boldsymbol{x}}^{3}(t) \oslash \dots \oslash \hat{\boldsymbol{x}}^{\text{F}}(t) \\
& \dots \\
\tilde{\boldsymbol{x}}^{\text{F}}(t) & = \boldsymbol{p} \oslash \hat{\boldsymbol{x}}^{1}(t) \oslash \hat{\boldsymbol{x}}^{2}(t) \oslash \dots \oslash \hat{\boldsymbol{x}}^{\text{F}-1}(t)
\end{split}
\end{equation}
Secondly, the similarity vector $\boldsymbol{\alpha^{\text{f}}}(t)$ is calculated using a matrix-vector multiplication (MVM). The similarity values are computed for each unbound estimate:
\begin{equation} \label{eq:res:FWD}
\begin{split}
\boldsymbol{\alpha}^{\text{f}}(t) & = \tilde{\boldsymbol{x}}^{\text{f}}(t) \cdot X^{\text{f}},\;\forall f \in [1,F]
\end{split}
\end{equation}
The estimates for the factors for the subsequent time step are given by the bipolarized, linear combination of all the codevectors with the similarity vectors acting as weights:
\begin{equation} \label{eq:BWD}
 \hat{\boldsymbol{x}}^{\text{f}}(t+1) = \mathrm{sign}(\boldsymbol{\alpha}^{\text{f}}(t) \cdot (X^{\text{f}})^T),\;\forall f \in [1,F]
\end{equation}
There is an alternative update approach: updating the factor's estimate sequentially. For every time step $t$, we start with updating the first factor's estimate $\hat{\boldsymbol{x}}^{\text{1}}(t+1)$ only. To update the second factor $\hat{\boldsymbol{x}}^{\text{2}}(t+1)$ we use the novel estimate of the first factor $\hat{\boldsymbol{x}}^{\text{1}}(t+1)$ and the latest estimates $\hat{\boldsymbol{x}}^{\text{f}}(t),\;\forall f \in [3,F]$. Equally, the update procedure continues until the last factor is updated always using the latest known estimate for each factor. We chose this sequential updating approach because it yields better performance for the resonator network.

The resonator network is said to be stuck in a limit cycle of length $l$ if all the estimated codevector reoccur after $l$ time steps but different codevectors occur in between as defined in equation~\eqref{eq:limitCylces}. In a deterministic implementation of the resonator network, there is no chance of breaking free of limit cycles resulting in non-convergence.
\begin{equation} \label{eq:limitCylces}
\hat{\boldsymbol{x}}^{\text{f}}(t+l) = \hat{\boldsymbol{x}}^{\text{f}}(t),\;\forall f \in [1,F]
\end{equation}

\clearpage
\subsection*{{Supplementary Note~2: Activation functions}}
The activation function is applied to the computed similarity values prior to the projection. A proper activation function would aim to \emph{separate} strong similarity values from the weak ones~\cite{universalHopfield}.
In the following, we discuss and elaborate on different types of activation functions.
\subsubsection*{Identity activation}
The neutral activation function is the identity function. There is no differentiation between strong and weak similarity values; the activation functions treat all the similarities equally. The identity function is used in the resonator networks~\cite{FradyResonator2020,KentResonatorNetworks2020} and is considered here as a baseline activation function. 

\subsubsection*{Sorting-based activation}
One possible implementation to separate the strong similarity values from the weak ones is to keep only the $K$ strongest similarity values. This activation function requires a sorting algorithm globally across all similarity values. Mathematically one can describe the sorting-based activation function as:
\begin{equation}\label{eq:topaP}
\alpha_{\text{i}}^{'} = S(\alpha)_{i} =\begin{cases}
    \alpha_{\text{i}}, & \text{if $\alpha_{\text{i}} \in \text{top}K(\alpha)$}\\
    0, & \text{otherwise},
  \end{cases}
\end{equation}
where $K$ is a tunable hyperparameter. We chose sorting-based activation to select only the strongest positive values. However, the sorting algorithm is costly for implementation in hardware.

\subsubsection*{Threshold-based activation}
To separate the strong similarity values from the weak ones, we propose a threshold-based activation function.
The proposed activation function applies a threshold $T$ on the similarity vector approximating a sorting-based $\text{top}K$. Following a winners-take-all approach, the activation function zeroes out the weak similarity values not reaching the given threshold. Similarity values that surpass the threshold are kept which yields a sparsely activated similarity vector:
\begin{equation}\label{eq:topaPT}
\alpha_{\text{i}}^{'} = S^{'}(\alpha)_{i} =\begin{cases}
    \alpha_{\text{i}}, & \text{if $\alpha_{\text{i}}>T$}\\
    0, & \text{otherwise}.
  \end{cases}
\end{equation}
This novel activation allows a local (as opposed to global sorting) and hardware-friendly implementation. The threshold is constant across iterations and dimensions.

On average, any given threshold activates a certain number of similarity values denoted by $K$. To map any desired number of activated values to a threshold $T$, we aim to find a function $T=f(K,D,M)$. We start by exploiting the randomness property of the high-dimensional vector space in which the similarity values are concentrated around a mean value. The distribution of the similarity values can be approximated by the normal distribution:
\begin{equation}\label{eq:normal}
\Pr(\alpha) = \frac{1}{{\sigma \sqrt {2\pi } }}e^{{{ - \left( {\alpha - \mu } \right)^2 }/{2\sigma ^2 }}},
\end{equation}
where $\mu=0$ and $\sigma=\frac{1}{\sqrt{D}}$.
To statistically activate a given number of $K$ similarity values, we first calculate the ratio between the similarity values to be activated and the total number of the similarity, given by the size of the corresponding codebook $M$:
\begin{equation}\label{eq:ratio}
r = K/M.
\end{equation}
To calculate a threshold such that on average only $K$ values exceed it, we invert the given ratio to find the ratio of deactivated similarity values:
\begin{equation}\label{eq:inverseRatio}
p = 1-r = 1 - K/M
\end{equation}
Using the quantile function $F_{T}$ (inverse probability density function) of the normal distribution we calculate a threshold such that on average only $K$ values exceed it.
In terms of the distribution function $F_{T}$, the quantile function returns the threshold $T$ such that:
\begin{equation}\label{eq:quantile}
F_{T}(t):=\Pr(T\leq t)=p
\end{equation}

Using Bayesian optimization we derived $K^{\star}$, the optimal number of the activated similarity values for each combination of $D$, $F$, and $M$. The optimal number $K^{\star}$ strongly depends on the number of the factors $F$ and weakly on the dimensions $D$, while it does not depend on the codebook size $M$. Fig.~\ref{fig:figsup2} shows the correlation between the optimal number of the activated values $K^{\star}$ and the corresponding threshold $T^{\star}$. For a fixed dimension $D$ and a number of the factors $F$, the optimal number of the activated values $K^{\star}$ remains constant for different sizes of the codebook $M$. This finding simplifies the optimal parameter search, as for a given dimension $D$ and a number of the factors $F$ only one single hyperparameter $K^{\star}$ has to be found. Knowing $K^{\star}$, $D$ and $M$ the optimal threshold can be calculated accordingly to the equations \eqref{eq:topaPT} to \eqref{eq:quantile} (see Table~\ref{tab:optthres}). For both threshold-based and sorting-based activations, we ended up with a comparable optimal number of the activated similarity values. Obviously, the sorting-based activation function is only capable of activating $K \in \mathbb{N}^{+}$ values, while the threshold-based activation function leads to $K \in \mathbb{R}^{+}$.

\begin{figure}[t]
\includegraphics[trim={0 22cm 0 0},clip,width=\textwidth]{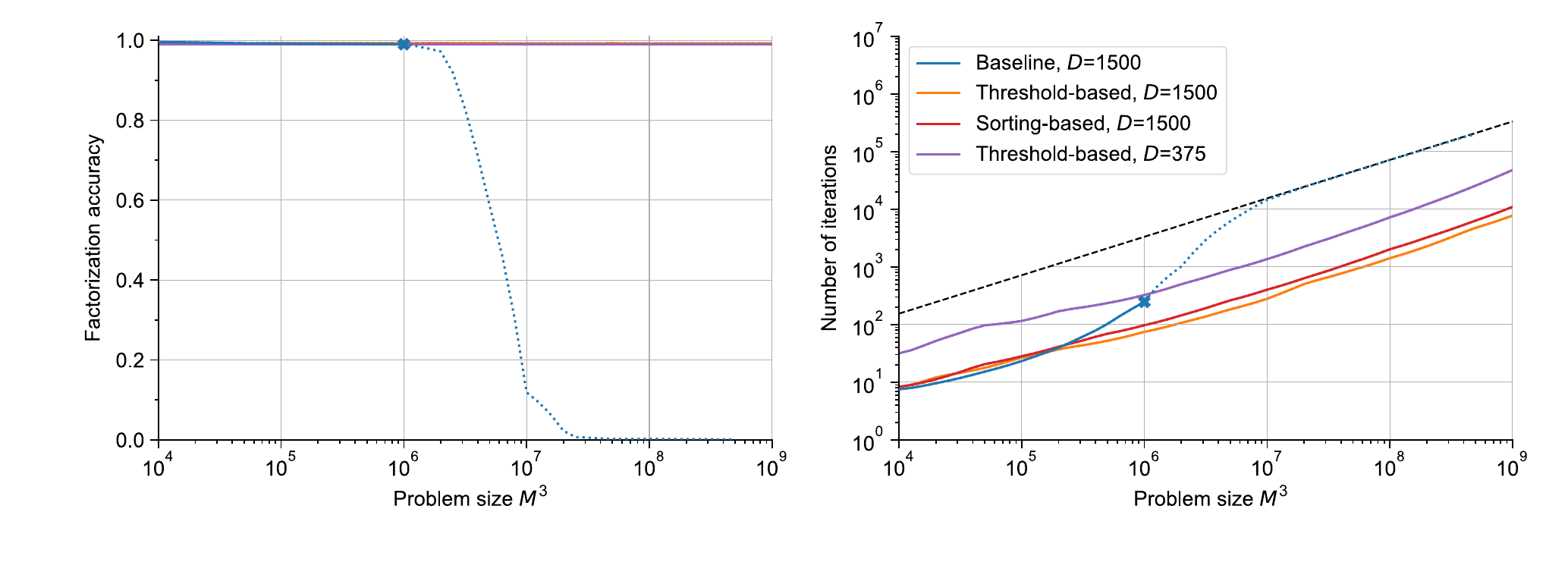}
\caption{
\textbf{Comparison of the sorting-based and the threshold-based activations as well as lowering dimensionality.}
The first line visualizes the threshold-based activation for $F=3$ at $D=1500$. The second line shows the sorting-based activation at the place of the threshold-based activation. The last line shows the threshold-based activation for a lower dimension $D=375$. The left side of the panel shows the accuracy for a wide range of problem sizes. It is clearly visible that the three aforementioned configurations reach an accuracy of 99\% and higher. The right subpanel shows the average number of iterations required to factorize a product vector for a wide range of problem sizes. The threshold-based and the sorting-based activations perform equally at the same dimensionality of $D=1,500$. Lowering the dimensionality by roughly four times requires more iterations to factorize the same product, resulting in a trade-off between dimensionality and time complexity. Compared to the baseline resonator, an in-memory factorizer with four times fewer dimensions is still capable of increasing the operational capacity.
}
\label{fig:figsup1}
\end{figure}

\begin{figure}[t]
\includegraphics[trim={0 9cm 0 0},clip,width=\textwidth]{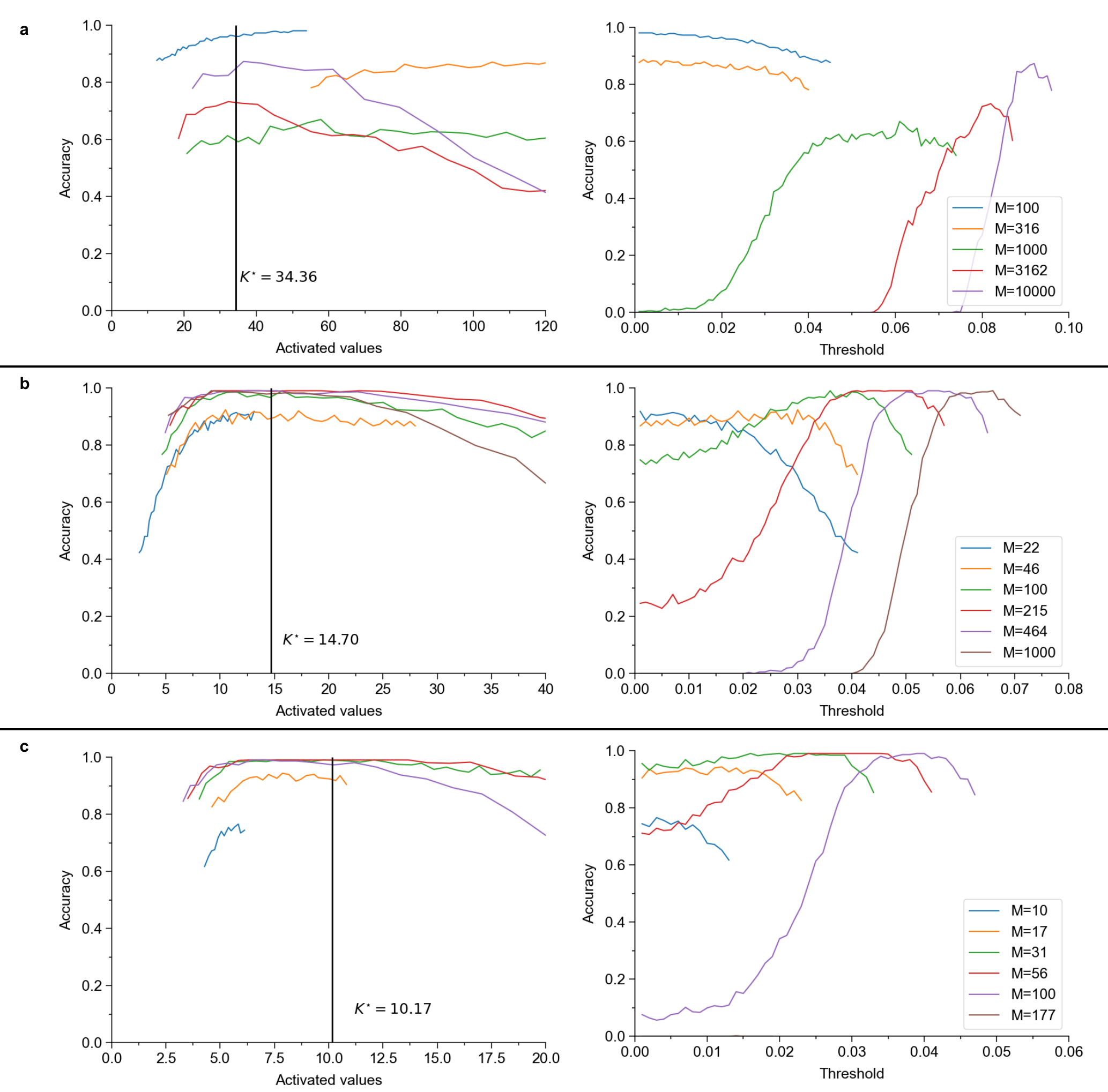}
\caption{
\textbf{Mapping of the optimal number of the top$K$-activated values ($\mathbf{K^{\star}}$) to the corresponding threshold ($\mathbf{T^{\star}}$).}
The left half of this figure shows the optimal number of the top$K$-activated values: \textbf{(a)} for $F=2$ and $D=1,000$;  \textbf{(b)} for $F=3$ and $D=1,500$; and \textbf{(c)} for $F=4$ and $D=2,000$.
The X-axis shows the number of the top$K$-activated values, and the Y-axis shows the accuracy of factorization.
We limited the number of the iterations to $N'=0.1 \times N$ 
to enforce the better solutions. Each line corresponds to a different codebook size such that the total problem size $M^{\text{F}}$ ranges from 
${1e4}$ to ${1e9}$. 
To get the curves displayed, we grid-searched the number of the activated values for each triplet $D$, $F$, and $M$. To convert the desired number of the activated values to a threshold, we use the approach as in the equations 
\ref{eq:topaPT} 
to \ref{eq:quantile}. 
The subpanels on the right side show the corresponding thresholds for the activated values curves. 
One can see that the best performing number of the top$K$-activated values is independent of $M$. 
The optimal number of the top$K$-activated values is averaged across all the optimas for each given codebook size.
}
\label{fig:figsup2}
\end{figure}

The sorting-based and the threshold-based activation functions perform equally as seen in Fig.~\ref{fig:figsup1}. Exploiting the threshold-based activation function, the dimensions of the in-memory factorizer can be reduced by at least four times without any decrease in the operational capacity.

\clearpage
\subsection*{{Supplementary Note~3: PCM noise model and its impact on factorization performance}} 
We present a detailed description PCM conductance variations by separating out the 
three key components, namely, programming noise, drift, and read noise.
\begin{equation}
\label{eq:noise_pcm}
\begin{split}
    G_\text{T0} & = G_\text{tar} + n_\text{p} \\
    G(t) & = G_\text{T0} \left(\frac{t}{T_0}\right)^{-\nu_{n_\nu}} + n_\text{r}, \\
\end{split}
\end{equation}
where $n_p\sim{\mathcal{N}(0,\sigma_p^2)}, n_\nu\sim{\mathcal{N}(0,\sigma_\nu^2)}, n_r\sim{\mathcal{N}(0,\sigma_r^2)}$ are random variables drawn from the normal distribution ${\mathcal{N}(\mu,\sigma^2)}$. $\sigma_p^2, \sigma_\nu^2, \sigma_r^2$ in the above equations represent the variance in programming noise, drift, and read noise, respectively. To estimate the variance parameters, we first program 65,536 devices on the experimental platform to the target conductance $G_\text{tar}$ of 5~$\mu$S. Then the conductance of all devices is measured roughly at 250\,s time intervals starting from initial time $T_0$ = 60\,s until 720,000\,s. The temporal conductance variation for each device is fitted to a function of form $\hat{G}(t) = G_{T0}(\frac{t}{T_0})^{-{\nu}}$ with measured $(t,G(t))$ pairs to find the device level estimates $\hat{G}(T_{0})$ and $\hat{\nu}$. The mean and the standard deviation of $\hat{\nu}$ are computed to derive the mean drift, ${\nu}$, and the drift variability, $\sigma_\nu^2$. $\sigma_\text{p}^2$ is derived by finding the variance of $G_{T0}$ distribution. $\sigma_\text{r}^2$ is determined using the residual distribution after deducting drifted conductance from the measured conductance $G(t)-\hat{G}(t)$. The derived parameters are as follows: $G_\text{tar} = 5~\mu\text{S}$, $T_0 = 60\text{s}$, $\sigma_\text{p} =  1.1636~\mu\text{S}$, $\sigma_\text{r} =  0.3951~\mu\text{S}$, $\sigma_{\nu} =  0.0907$ and $\nu = 0.0428$.

To separately quantify the effect of read and programming noise on the same configuration of the in-memory factorizer used in Section II, we conduct simulations in which we sweep the standard deviation of (i) the  programming noise and (ii) the read noise while keeping the rest of the PCM model parameters at default values. The results are shown in Fig.~\ref{fig:prog_read_noise}. The green vertical line in each figure corresponds to the default standard deviation in the respective noise component observed on the experimental platform. With the insights obtained from the figures, we can assess how well the in-memory factorizer performs in the presence of both read and programming noise. In the case of read noise, the factorizer is robust up to a standard deviation of 0.921\,$\mu$S, whereas the corresponding value for the standard deviation of the programming noise is 1.629\,$\mu$S. Both these values are well above the observed level of noise in the experimental platform, which is 0.3951\,$\mu$S and 1.1636\,$\mu$S respectively for read and programming noise.

Next, we evaluate the sensitivity of the factorizer performance to the total crossbar noise. For this, starting from zero standard deviation for programming and read noise, we gradually increase both noise levels while maintaining the ratio of standard deviation between read noise to programming noise as observed on the experimental platform $(\sigma_\text{r}/\sigma_\text{p}=0.3951/1.1636)$. The results are given in Extended Fig 1.
We observe that at zero noise, the factorizer performs poorly with an accuracy of 25.4\% while requiring 16,000 iterations to converge on average, due to the deterministic nature of the search hindering the possibility to break the limit cycles. The factorizer operates at its peak performance when the standard deviation of the total noise is maintained within the range [0.293~$\mu$S, 1.277~$\mu$S]. The standard deviation of the total noise observed on the experimental platform (0.98~$\mu$S) falls safely in the middle of this tolerated noise range. 

Further, in Fig.~\ref{fig:noise_source}, we simulate the performance of the in-memory factorizer when the noisy device conductances on the forward path (similarity calculation) and the backward path (projection) are identical versus sampled separately from the noise distribution. The first case represents the operations performed on a single core whereas the second case represents operations on two different cores. The peak performance in each case is approximately the same, with operations on the same core having a marginal improvement in the degradation of performance at higher noise levels.

\clearpage
\subsection*{{Supplementary Note~4: Hardware design}} 
Here, we provide an analysis and comparison of hardware designs dedicated to the energy-efficient execution of the in-memory factorizer. The proposed in-memory factorizer performs the dominant similarity calculation and projection on the next generation in-memory crossbar arrays together with analog peripherals such as input pulse width modulation (PWM) circuits (for digital-to-analog (DAC) conversion) and analog-to-digital conversion (ADC) circuits. The remaining operations including unbinding, permutations, and the activation function are implemented on custom-designed peripheral digital processing units. We compare the proposed in-memory factorizer's hardware design with a dedicated reference design comprising digital CMOS multipliers and accumulators to perform MVM operations for similarity calculation and projection. To implement the remaining operations in the reference design, the custom peripheral digital processing units designed for the in-memory factorizer are reused. 

To get a performance estimation of the respective designs, we designed all the digital logic modules using \textit{SystemVerilog} and synthesized them with Samsung's 14\,nm LPP technology node using Cadence's \textit{Genus} synthesizer tool. The digital circuits are run at a frequency of 10\,MHz, yielding a clock period of 100\,ns. This allows the peripheral digital circuits to synchronize with the analog crossbar arrays which operate with approximately the same delay. For the power estimation, we used the typical corner case which applies a supply voltage of 0.8\,V and assumes a temperature of $25^\circ$C. All design parameters and operating conditions corresponding to the digital circuits are presented in Table~\ref{tab:cct_cond}.  The switching activity of the digital netlist is derived via post-synthesis simulation.

To estimate the energy consumed by the analog/mixed-signal component of the in-memory factorizer design, we consider the design parameters given in Table~\ref{tab:cct_cond} and make reasonable projections on the analog crossbar designs discussed in \cite{mixedprec,9508706}. The energy consumed by the analog component comprises five parts: 
\begin{itemize}
    \item The energy for PWM generation: This energy is calculated as a function of the mean power dissipated by the circuits that drive the PWM pulses to the source lines, the active time of PWM, and the number of active source lines.
    
    \item The energy for charging and discharging source line capacitance: This energy is calculated as a function of source line capacitance, supply voltage, and the number of active rows, and columns on the crossbar.
    
    \item The energy dissipated on the crossbar: This energy is calculated as a function of read voltage, the number of active rows and columns, average device conductance, supply voltage, and active time of PWM.
    
    \item The energy for biasing the regulators at the end of bit lines: This energy is calculated as a function of regulator bias current, supply voltage, maximum PWM active time, and the number of active columns(bit-lines) on the crossbar.
    
    \item The energy for analog-to-digital conversion of the output: This energy is calculated as a function of energy per conversion step of a current-controlled oscillator-based ADC~\cite{9508706}, output bit precision, and the number of active columns on the crossbar.
\end{itemize}

The main functional difference between the proposed in-memory factorizer design and the reference digital design is that the reference design's MVM output does not include any stochastic noise sources whereas the proposed in-memory factorizer inherently does. For this reason, for a given dimension $D$ and problem size $M$, the two designs demonstrate different average accuracy and different average number of iterations. For example, at $D=256$, $M=256$, the proposed in-memory factorizer design converges to correct results with 99.74\% accuracy using 3058 iterations on average (i.e., within the operational capacity), whereas the reference digital design only achieves 95.76\% accuracy using on average 3802 iterations. Hence, this combination of parameters ($D=256$, $M=256$) does not lead to the same operational capacity and hence it is not appropriate to compare the two designs for these combinations of parameters.

Instead, we experiment with increasing the dimensionality of the reference design from $D=256$ until it reaches $\geq 99\%$ accuracy and the same number of iterations to converge compared to the proposed in-memory factorizer. We find that the dimension that achieves this behavior in the reference design is $D=352$. This intermediate comparison point also lacks fairness because of the imbalance in the number of MAC operations consumed between the reference design and the proposed design.

Therefore we consider the final comparison point that involves larger dimensions and problem sizes. This is motivated by two factors. 
\begin{enumerate}
    \item With an increasing number of dimensions, the reference design with deterministic behavior tends to achieve higher accuracies and a lower average number of iterations, even when the problem size is increased at the same rate. This allows the comparison of hardware built with the same parameters which achieves the same algorithmic performance.
    
    \item We have observed that the PCM crossbar array sizes can be scaled up to 512x512 size~\cite{Y2021narayananVLSI} without severely compromising the signal-to-noise ratio. Furthermore, the energy of the analog core comprising the crossbar array is dominated by the peripheral circuitry whose energy scales linearly with the size of the crossbar. In comparison, the energy scales quadratically in the reference digital design. 
\end{enumerate}

\begin{figure}[t]
\includegraphics[width=0.8\textwidth]{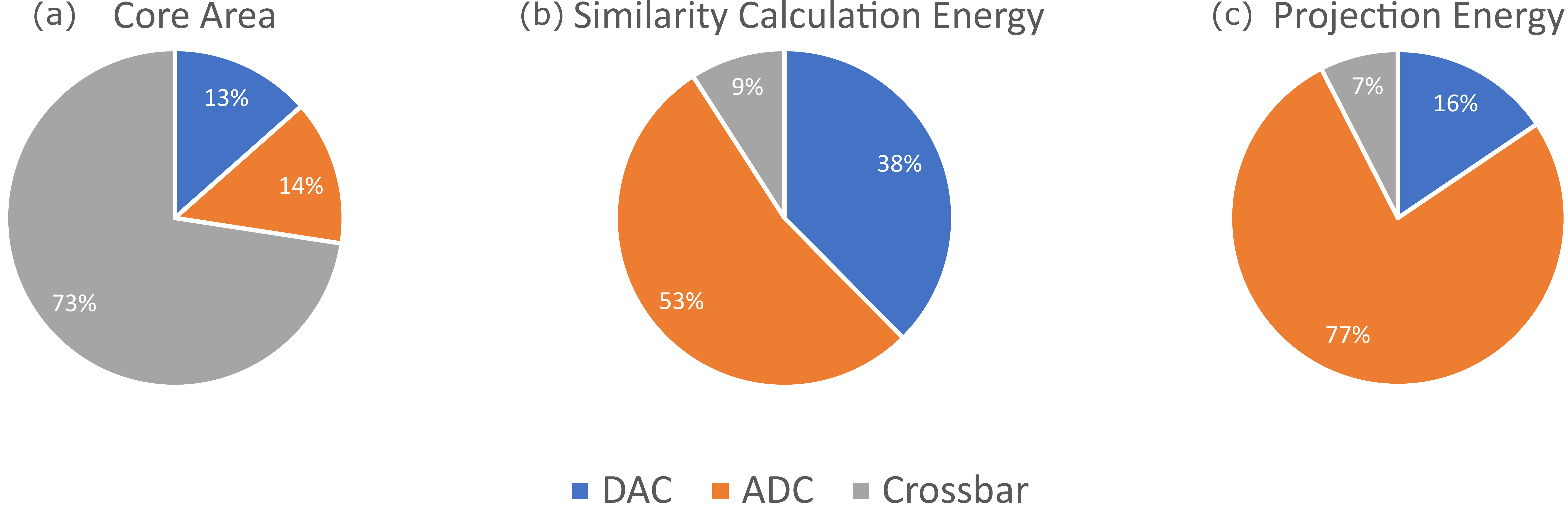}
\caption{
\textbf{(a) Area and (b), (c) energy breakdown of the proposed in-memory factorizer.}
The total tile area of 551,000~$\mu$m$^2$ breaks down into identical fractions for similarity calculation and projection because of the use of the same hardware. The energy breakdown is however different due to several reasons such as the use of different input/output data precision and different ADC integration times. Here, the energy consumed by biasing the regulators and charging and discharging of the bit line capacitances as discussed in Supplementary Note~3, are grouped under the overall ADC energy. DAC energy and area are those attributed to the PWM generation circuits. The total energy per iteration is 2.87~nJ and 1.94~nJ for similarity calculation and projection, respectively.
}
\label{fig:figsup3}
\end{figure}

\begin{figure}[tbh]
 \includegraphics[width=\textwidth]{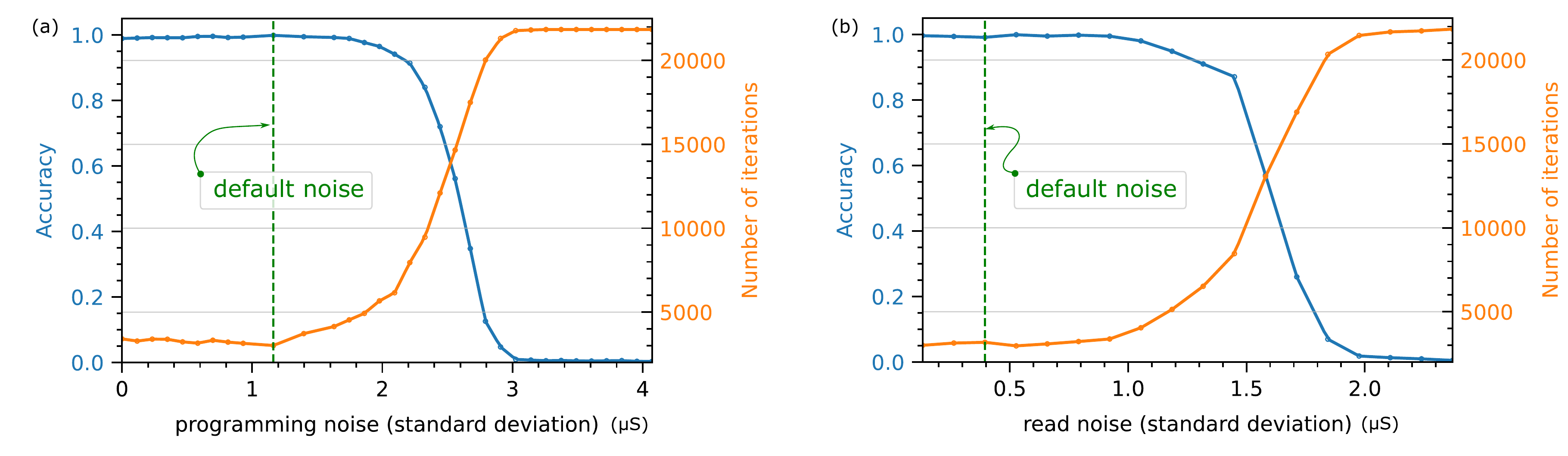}
 \caption{
 \textbf{Performance in the presence of noise.} The factorization accuracy and the number of iterations required for convergence as a function of (a)~the standard deviation of the programming noise, (b)~the standard deviation of the read noise while the remaining noise parameters are fixed to the default values as listed in Supplementary Note~3. The green vertical line corresponds to the default noise level in the respective component.
 }
\label{fig:prog_read_noise}
\end{figure}

\begin{figure}[tbh]
 \includegraphics[width=0.5\textwidth]{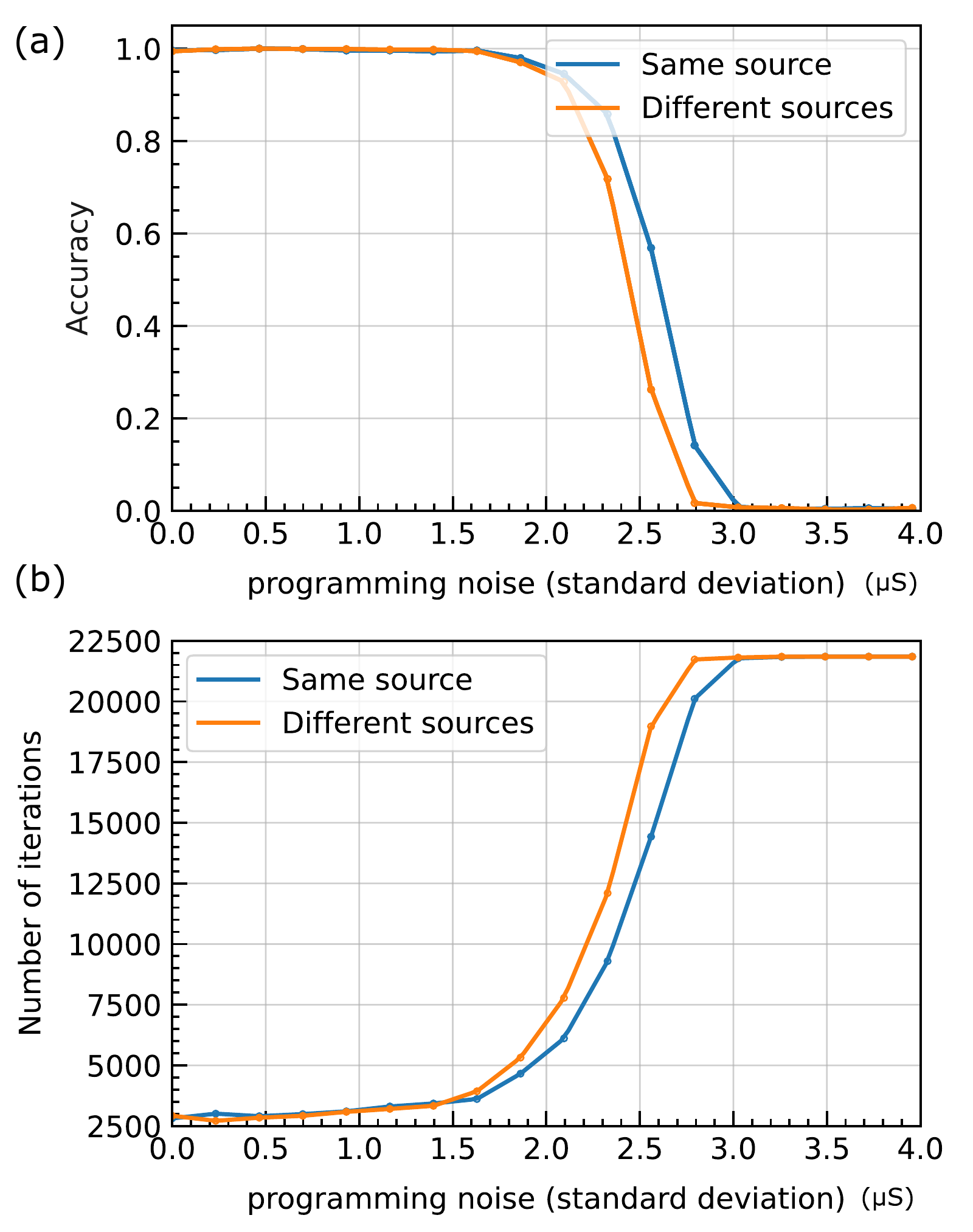}
 \caption{
 \textbf{Performance as a function of the noise source.} (a)~Factorization accuracy and (b)~the number of iterations required for convergence as a function of the standard deviation of programming noise. In the \emph{same source} case, the noise is sampled once for both the forward path (similarity calculation) and the backward path (projection) while in \emph{different sources} case it is sampled separately for the two operations. The remaining noise parameters are fixed to the default values as listed in Supplementary Note~3.
 }
\label{fig:noise_source}
\end{figure}

We choose $D=512$ and $M=512$ as the parameter combination for the final comparison point. Our simulations find that with these parameter combinations, the reference design and the proposed design achieve 100\% accuracy with 6554 iterations and 6185 iterations on average respectively to factorize a query. The proposed design consumes 5.35\,nJ to complete a single iteration, with a breakdown of 1.94\,nJ, 2.49\,nJ, and 0.5\,nJ shares for projection, similarity calculation, and the rest of the peripheral operations respectively. Further energy breakdown of projection and similarity calculation in terms of the digital-to-analog conversion, the crossbar operation, and the analog-to-digital conversion is presented in Fig.~\ref{fig:figsup3}. The main reasons for the energy difference in the two cases are:
\begin{itemize}
  \setlength{\itemsep}{0pt}
  \setlength{\parskip}{0pt}
    \item The number of active inputs: in the case of similarity calculation 100\%, in the case of projection 7\%
    \item Integration time: 10~ns for similarity calculation and 40~ns for projection
    \item 8~bits ADC bit precision for similarity calculation whereas 1~bit for projection
\end{itemize}

The reference design with the parameters $D=512$, $M=256$ was found to be consuming 61.4\,nJ energy to complete a single iteration. This can be split as 46.3\,nJ, 14.5\,nJ, and 0.5\,nJ shares for the above-mentioned operational stages. Table~\ref{tab:AnalogvsDigital} and Fig.~\ref{fig:figsup3} summarizes all results. The total energy consumption for a single iteration as well as the average energy consumption for processing a single query until its factorization is complete in both designs are listed in the table. It can be seen that our proposed design outperforms the reference design by a factor of 11.5$\times$, 12.2$\times$, and 4.85$\times$ when iteration energy, average query factorizing energy, and area savings are considered respectively in the final comparison point thanks to higher energy efficiency obtained when scaling up the MVM operation using in-memory computing approach.

\clearpage

\clearpage

\clearpage

\clearpage

\clearpage

\clearpage
\begin{table}[]
\caption{The optimal number of activated values corresponding to the combination of number of factors and vector dimensions.}
\label{tab:optthres}
\begin{tabular}{l|llll|llll|llll}
Factors ($F$)    & \multicolumn{4}{c|}{2}  & \multicolumn{4}{c|}{3}       & \multicolumn{4}{c}{4}     \\ \hline
Dimensions ($D$) & 256 & 512 & 1024 & 2048 & 256  & 512   & 1024  & 2048  & 256  & 512  & 1024 & 2048 \\ \hline
Activated Values  & 20.79 &  39.98 & 54.79 & 104.87 & 8.34 & 10.30 & 11.02 & 13.60 & 5.81 & 6.23 & 6.87 & 8.13
\end{tabular}
\end{table}

\clearpage
\begin{table}[h]
\centering
\caption{Design parameters and operating conditions.}
\label{tab:cct_cond}
\begin{tabular}{lll}
\textbf{}                & \textbf{Symbol}     & \textbf{Value}    \\
\hline
\textbf{Analog design component}               
\\ \hline

Read voltage            & $V_{read}$ & 0.1\,V    \\
Maximum conductance         & $G_{max}$  & 10\,$\mu S$ \\
Source line capacitance & $C_{SL}$   & 4\,fF      \\
Regulator bias current  & $I_{bias}$ & 50\,$\mu A$ \\
PWM circuit power       & $P_{FSM}$  & 54\,$\mu W$   \\
ADC conversion energy   & $E_{ADC}$  & 5.0\,fJ/conv-step    \\
Unit cell composition    & - & 2T2R    \\
\\ \hline
\textbf{Digital design component}   
\\\hline
Technology node           & -     & 14\,nm    \\
Supply voltage            & $VDD$ & 0.8\,V    \\
Operating frequency       & $F$ & 10\,MHz    \\
Corner                    & -   & Typical    \\
Temperature               & -   & 25\,\textdegree C   \\ \hline
\end{tabular}
\end{table}

\clearpage
\begin{table}[]
\caption{Comparison between the proposed in-memory factorizer and an equivalent reference digital design.}
\begin{tabular}{lccl}
\hline
                                                                & Reference  & \textbf{Proposed} & Unit  \\
\textbf{Projection }                                      &        &                      \\ \hline
\multicolumn{1}{r}{Area }                                       & 4.6    &   0.551    & mm$^2$     \\
\multicolumn{1}{r}{Time per iteration }                         & 100    &   40       & ns     \\
\multicolumn{1}{r}{Energy per iteration}                        & 46.3   &   1.94     & nJ     \\
\multicolumn{1}{r}{Peak throughput}                             & 5.24   &   13.1     & TOPS   \\
\multicolumn{1}{r}{Energy efficiency}                           & 11.3   &   270      & TOPS/W \\
\multicolumn{1}{r}{Area efficiency}                             & 1.15   &   23.8     & TOPS/mm$^2$ \\\\ 
\textbf{Similarity calculation}                                       &        &                      \\ \hline
\multicolumn{1}{r}{Area }                                       & 0.865  &   0.551     & mm$^2$     \\
\multicolumn{1}{r}{Time per iteration }                         & 100    &   10        & ns     \\
\multicolumn{1}{r}{Energy per iteration}                        & 14.5   &   2.87      & nJ     \\
\multicolumn{1}{r}{Peak throughput}                             & 5.24   &   52.4      & TOPS   \\
\multicolumn{1}{r}{Energy efficiency}                           & 36.1   &   182       & TOPS/W \\
\multicolumn{1}{r}{Area efficiency}                             & 6.06   &   95.1      & TOPS/mm$^2$ \\\\ 
\textbf{Other peripherals}                                               &        &                   \\ \hline
\multicolumn{1}{r}{Area }                                       & 24.3   &   24.3      & $\mu$m$^2$     \\
\multicolumn{1}{r}{Time per iteration }                         & 300    &   300       & ns     \\
\multicolumn{1}{r}{Energy per iteration}                        & 0.53   &  0.53       & nJ     \\\\  
\textbf{Total}                                                          &        &                   \\ \hline
\multicolumn{1}{r}{Area }                                       & 5.46    &   1.13       & mm$^2$     \\
\multicolumn{1}{r}{Time per iteration }                         & 500    &   350       & ns     \\
\multicolumn{1}{r}{Energy per iteration}                        & 61.4   &   5.35      & nJ     \\
\multicolumn{1}{r}{Average iterations per factorization }       & 6553   &   6184      & -     \\
\multicolumn{1}{r}{Time per factorization}                     & 3280   &   2164      & us     \\
\multicolumn{1}{r}{Energy per factorization}                    & 402   &    33.1      & $\mu$J     \\ \hline
\end{tabular}
\label{tab:AnalogvsDigital}
\end{table}

\clearpage

\end{document}